\documentclass[eqsecnum,noshowpacs,showkeys,nofootinbib,aps]{revtex4}

\usepackage{epsfig}
\usepackage{xcolor}
\usepackage{ulem} 

\renewcommand{\theequation}{\arabic{equation}}
\def\be{\begin{equation}}
\def\ee{\end{equation}}
\def\bea{\begin{eqnarray}}
\def\eea{\end{eqnarray}}

\def\na{\nabla}

\begin{document}

\title{GEMS embeddings of Schwarzschild and RN black holes\\
in Painlev\'e-Gullstrand spacetimes}
\author{Soon-Tae Hong}
\email{soonhong@sogang.ac.kr} \affiliation{Center for Quantum
Spacetime and
\\ Department of Physics, Sogang University, Seoul 04107, Korea}
\author{Yong-Wan Kim}
\email{ywkim65@gmail.com}
 \affiliation{Department of Physics and
 \\ Research Institute of Physics and Chemistry, Jeonbuk National University, Jeonju 54896, Korea}
\author{Young-Jai Park}
\email{yjpark@sogang.ac.kr} \affiliation{Center for Quantum
Spacetime and
\\ Department of Physics, Sogang University, Seoul 04107, Korea }
\date{\today}

\begin{abstract}
Making use of the higher dimensional global embedding Minkowski
spacetime (GEMS), we embed (3+1)-dimensional Schwarzschild and
Reissner-Nordstr\"om (RN) black holes written by the
Painlev\'e-Gullstrand (PG) spacetimes, which have off-diagonal
components in metrics, into (5+1)- and (5+2)-dimensional flat
ones, respectively. As a result, we have shown the equivalence of
the GEMS embeddings of the spacetimes with the diagonal and
off-diagonal terms in metrics. Moreover, with the aid of their
geodesic equations satisfying various boundary conditions in the
flat embedded spacetimes, we directly obtain freely falling
temperatures. We also show that freely falling temperatures in the
PG spacetimes are well-defined beyond the event horizons, while
they are equivalent to the Hawking temperatures, which are
obtained in the original curved ones in the ranges between the
horizon and the infinity. These will be helpful to study GEMS
embeddings of more realistic Kerr, or rotating BTZ black holes.
\end{abstract}


\maketitle

\section{introduction}
\setcounter{equation}{0}
\renewcommand{\theequation}{\arabic{section}.\arabic{equation}}

Any low dimensional Riemannian manifold can be locally
isometrically embedded in a higher dimensional flat one
\cite{Fronsdal:1959zza,Rosen:1965,Goenner1980}. To be specific, by
using isometric embedding, it is expected that it may be possible
to find the higher dimensional flat manifold than the original
curved one with singularity. Moreover, this global embedding
Minkowski spacetime (GEMS) method can be used to show their
equivalence of Hawking \cite{Hawking:1974sw} and Unruh effects
\cite{Unruh:1976db}. Deser and Levin
\cite{Deser:1997ri,Deser:1998bb,Deser:1998xb} showed that the
Hawking temperature for a fiducial observer in a curved spacetime
can be considered as the Unruh one for a uniformly accelerated
observer in a higher-dimensional flat spacetime. Since then, there
have been a lot of works on the GEMS approach to confirm these
ideas in various other spacetimes
\cite{Hong:2000kn,Kim:2000ct,Hong:2003xz,Chen:2004qw,Santos:2004ws,
Banerjee:2010ma,Cai:2010bv,Hu:2011yx,hong2000,hong2001,
hong2005,hong2004,hong2006,Paston:2013mfa,paston2014,Sheykin:2019uwj}
and an interesting extension to embedding gravity
\cite{Paston:2011wp,Paston:2018orc,Paston:2020otr,Paston:2020isf}.
Later, Brynjolfsson and Thorlacius (BT) \cite{Brynjolfsson:2008uc}
introduced a local temperature measured by a freely falling
observer in the GEMS method. The local free fall temperature they
obtained remains finite at the event horizon and it approaches the
Hawking temperature at spatial infinity. Here, a freely falling
local temperature is defined at special turning points of radial
geodesics where a freely falling observer is momentarily at rest
with respect to a black hole. Following the BT's approach, we have
extended the methods to various interesting curved spacetimes
\cite{Kim:2009ha,Kim:2013wpa,Kim:2015wwa,hong2015,Hong:2018spz,Hong:2019zsi}
to investigate local temperatures of corresponding spacetimes and
their equivalence to Hawking ones. Meanwhile, radial geodesics can
be categorized into drip, rain, and hail frames according to which
initial states observers or objects are following
\cite{Tayler:2000}. Observers are said to be in a drip frame when
they are freely falling from rest at a finite initial radius
$r_0$. In particular, when observers are freely falling at rest
from infinitely far away or $r_0\rightarrow\infty$, they are said
to be in a rain frame. Observers in a hail frame are hurled inward
with initial velocity $v_\infty$ toward black holes at infinity.
In this classification scheme, freely falling local temperature of
the BT's method is the one seen by an observer in a drip frame. It
is interesting to extend the BT's work to the other two frames.

On the other hand, the PG coordinates
\cite{Painleve,Gullstrand:1922tfa} have been newly recognized as
an alternative to the Schwarzschild coordinates to avoid the
singularity at the event horizon. The PG spacetime remains regular
across the horizon. It describes stationary, but not static
spacetimes. Observers falling into a black hole in the PG
spacetime use their own proper times with spatial flat
hypersurfaces flowing radially inward. The importance of the PG
spacetime was newly received attention related to the Hawking
radiation as a tunnelling process
\cite{Kraus:1994fh,Parikh:1999mf,Parikh:2004ih}. In this scenario,
when self-gravitation of particles is taken into account, positive
energy virtual particles triggered by vacuum fluctuations just
inside an event horizon can tunnel out across the horizon. It was
extended to more general spherical spacetimes with interesting
physical interpretations
\cite{Martel:2000rn,Francis:2003rj,Hamilton:2004au,Finch:2012fq}.
Laboratory analogues of gravity
\cite{Unruh:1980cg,Visser:1997ux,Volovik:2003,Barcelo:2005fc} can
also be recast in the PG spacetime coordinates by conformal
transformations. Moreover, the PG spacetime used to describe
gravitational collapse dynamically both inside and outside the
horizon in a single coordinate patch \cite{Kanai:2010ae}, compared
to the standard model of Oppenheimer and Snyder
\cite{Oppenheimer:1939,MTW:1973} which uses two different
spacetimes corresponding to the interior and exterior regions of
the collapsing body. From a modern point of view, several of the
subtle features of the PG spacetime are under investigation
\cite{Visser:2003tt,Perez-Roman:2018hfy,Faraoni:2020ehi,Baines:2020egv}.

The main goal of this paper is the construction and analysis of
the GEMS embeddings of spacetimes with off-diagonal components in
metric, whose embedding is extremely difficult to carry out, and
actually there are no such GEMS embedding models as far as we
know. Since the more physically realistic Kerr, or rotating BTZ
holes have the off-diagonal terms in metrics, it would be
interesting enough in itself if their GEMS embeddings can be
found. If then, one can further find various thermodynamic
functions including freely falling temperatures seen by a freely
falling observer beyond event horizons. As a preliminary to such a
road, in this paper, we will study first GEMS embeddings of the
Schwarzschild and the Reissner-Nordstr\"om (RN) black holes
described by the PG spacetimes, which are also known to be
nontrivial due to the existence of off-diagonal terms in metrics.
We will also show that their Hawking and Unruh temperatures are
the exactly same with the ones in the original GEMS approach.
Moreover, by following BT's method and full geodesic equations
satisfying various boundary conditions, we will find freely
falling temperatures seen by observers in the PG-embedded
Schwarzschild and RN black holes, which can be extended smoothly
through the future event horizon. The remainder of the paper is
organized as follows. In Section II, we will construct GEMS
embeddings and various temperatures of the Schwarzschild black
hole in the PG spacetime and compare them with the ones of the
Schwarzschild black hole in the original spherically symmetric
spacetime. In Section. III, we also present GEMS embeddings of the
RN black hole in the PG spacetime and study various temperatures
in drip, rain and hail frames. Conclusions are drawn in Section
IV.

\section{GEMS embedding of the Schwarzschild black hole in the PG spacetime}
\setcounter{equation}{0}
\renewcommand{\theequation}{\arabic{section}.\arabic{equation}}

\subsection{GEMS embedding of the Schwarzschild black hole in the spherically symmetric spacetime}

In this subsection, we briefly recapitulate the GEMS embeddings
\cite{Fronsdal:1959zza,Rosen:1965,Goenner1980} and their way of
finding Hawking, Unruh, freely falling temperatures at rest
\cite{Brynjolfsson:2008uc} of the Schwarzschild black hole in the
spherically symmetric spacetime. The spherically symmetric
Schwarzschild black hole is described by the metric
 \be\label{metric}
 ds^2=-f(r)dt^2+f^{-1}(r)dr^2+r^2(d\theta^2+\sin^2\theta d\phi^2)
 \ee
with
 \be\label{fr}
 f(r)=1-\frac{2m}{r}.
 \ee
From the metric, one can find the surface gravity \cite{wald} as
 \be\label{sgravity}
 k_H = \left.\sqrt{-\frac{1}{2}(\na^{\mu}\xi^{\nu})(\na_{\mu}\xi_{\nu})}~\right|_{r= r_H}
     =\frac{1}{2r_H},
 \ee
where $\xi^\mu$ is a Killing vector and the event horizon is given
by $r_H=2m$. Then, the Hawking temperature $T_H$ seen by an
asymptotic observer is given by
 \be\label{HawkingT}
 T_H=\frac{k_H}{2\pi}=\frac{1}{4\pi r_H}.
 \ee
Moreover, a local fiducial temperature measured by an observer who
rests at a distance from the black hole is given by
 \be\label{fidT}
 T_{\rm FID}(r)=\frac{T_H}{\sqrt{f(r)}}=\frac{r^{1/2}}{4\pi r_H(r-r_H)^{1/2}}.
 \ee
Note that the fiducial temperature $T_{\rm FID}$ diverges at the
event horizon, while it becomes the Hawking temperature to an
asymptotic observer.

Now, by following the GEMS approach
\cite{Fronsdal:1959zza,Rosen:1965,Goenner1980}, the
(3+1)-dimensional spherically symmetric Schwarzschild spacetime
(\ref{metric}) can be embedded in a (5+1)-dimensional flat one as
 \be
 ds^{2}= \eta_{IJ}dz^Idz^J,~{\rm with}~\eta_{IJ}={\rm
 diag}(-1,1,1,1,1,1),
 \ee
where embedding coordinates are given by
 \bea\label{gems-sch}
 z^{0}&=&k_{H}^{-1}f^{1/2}(r)\sinh k_{H}t, \nonumber \\
 z^{1}&=&k_{H}^{-1}f^{1/2}(r)\cosh k_{H}t, \nonumber \\
 z^{2}&=&r\sin\theta\cos\phi, \nonumber \\
 z^{3}&=&r\sin\theta\sin\phi, \nonumber \\
 z^{4}&=&r\cos\theta, \nonumber \\
 z^{5}&=&\int dr
 \left(\frac{r_H(r^2+rr_H+r^2_H)}{r^3}\right)^{1/2}.
 \eea
Note the analyticity of $z^{5}$ in $r>0$. This helps the embedding
coordinates (\ref{gems-sch}) to cover the region $0<r<r_H$ by
analytic extension.

In static detectors ($r,~\theta,~\phi={\rm constant}$) described
by a fixed point in the ($z^2,~z^3,~z^4,~z^5$) plane on the GEMS
embedded spacetime, an observer who is uniformly accelerated in
the (5+1)-dimensional flat spacetime, follows a hyperbolic
trajectory in ($z^0,~z^1$) described by
 \be\label{acc6sch}
 a^{-2}_6=(z^1)^2-(z^0)^2= \frac{f(r)}{k^2_H}.
 \ee
Thus, one can arrive at the Unruh temperature for the uniformly
accelerated observer in the (5+1)-dimensional flat spacetime
 \be\label{unruh-sch}
 T_U=\frac{a_6}{2\pi}=\frac{r^{1/2}}{4\pi r_H(r-r_H)^{1/2}}.
 \ee
This corresponds to the fiducial temperature (\ref{fidT}) for the
observer located at a distance from the Schwarzschild black hole.
The Hawking temperature $T_H$ seen by an asymptotic observer can
be obtained as
 \be
 T_H=\sqrt{-g_{00}}T_U=\frac{k_H}{2\pi}.
 \ee
As a result, one can see that the Hawking effect for a fiducial
observer in a black hole spacetime is equal to the Unruh effect
for a uniformly accelerated observer in a higher-dimensional flat
spacetime.

It is appropriate to comment that the static detectors following a
timelike Killing vector $\xi=\partial_t$ with the same condition
($r,~\theta,~\phi={\rm constant}$) in the original
(3+1)-dimensional spacetime have constant 4-acceleration as
 \be\label{a4sch}
 a_4=\frac{r_H}{2r^{3/2}(r-r_H)^{1/2}},
 \ee
where the constant 4-acceleration is given by $a^2_4=a_\mu a^\mu$
with $a_\mu=\xi_{\nu;\mu}\xi^\nu/|\xi|^2$. Comparing this with the
acceleration (\ref{unruh-sch}) in the (5+1)-dimensional embedded
spacetime, one can have
 \be\label{a6a4}
 a_6=\frac{r^2}{r^2_H}a_4.
 \ee

On the other hand, before finding freely falling accelerations in
the embedded spacetime, let us first consider the geodesic
equation \cite{wald}
 \be\label{gdeq}
 \frac{d^2x^\mu}{d\tau^2}+\Gamma^\mu_{\nu\rho}\frac{dx^\nu}{d\tau}\frac{dx^\rho}{d\tau}=0,
 \ee
where $x^\mu=(t,r,\theta,\phi)$. Here, the Christoffel symbol is
given by
 \be
 \Gamma^\rho_{\mu\nu}=\frac{1}{2}g^{\rho\sigma}
              (\partial_\mu g_{\nu\sigma}+\partial_\nu g_{\mu\sigma}-\partial_\sigma g_{\mu\nu}).
 \ee
From the metric (2.1),
the geodesic equation gives us explicitly as follows
 \bea
 &&\frac{dv^0}{d\tau}+\frac{2m}{r^2-2mr}v^0v^1=0, \\
 &&\frac{dv^1}{d\tau}+\frac{m}{r^3}(r-2m)(v^0)^2
       -\frac{m}{r^2-2mr}(v^1)^2-(r-2m)\left[(v^2)^2+\sin^2\theta(v^3)^2\right]=0,     \\
 &&\frac{dv^2}{d\tau}+\frac{2}{r}v^1v^2-\sin\theta\cos\theta(v^3)^2=0, \\
 &&\frac{dv^3}{d\tau}+\frac{2}{r}v^1v^3+2\cot\theta v^2 v^3=0,
 \eea
where $v^\mu=dx^\mu/d\tau$ denotes the four velocity vector.
Without loss of generality, one can consider the geodesics on the
equatorial plane given by $\theta=\pi/2$. Then, one has
$v^2=d\theta/d\tau=0$ and the geodesic equations are reduced to
 \bea
 &&\frac{dv^0}{d\tau}+\frac{2m}{r^2-2mr}v^0v^1=0, \label{ge0}\\
 &&\frac{dv^1}{d\tau}+\frac{m}{r^3}(r-2m)(v^0)^2
       -\frac{m}{r^2-2mr}(v^1)^2-(r-2m)(v^3)^2=0,     \\
 &&\frac{dv^3}{d\tau}+\frac{2}{r}v^1v^3=0.\label{ge2}
 \eea
One can easily integrate Eqs. (\ref{ge0}) and (\ref{ge2}) as
 \bea
 v^0&=&\frac{dt}{d\tau}=\frac{c_1 r}{r-2m}, \label{v0s}\\
 v^3&=&\frac{d\phi}{d\tau}=\frac{c_2}{r^2}, \label{v2s}
 \eea
respectively, where $c_1$ and $c_2$ are integration constants
\cite{Hong:2020bdb}.

Note that for the Killing vectors $\xi^\mu=(1,0,0,0)$ and
$\psi^\mu=(0,0,0,1)$, two conserved quantities of $E$ and $L$ are
given by
 \bea
 E &=& -g_{\mu\nu}\xi^\mu v^\nu =\left(1-\frac{2m}{r}\right) v^0,\label{K1}\\
 L &=& g_{\mu\nu}\psi^\mu v^\nu =r^2 v^3.\label{K2}
 \eea
Comparing these relations with Eqs. (\ref{v0s}) and (\ref{v2s}),
one can fix the integration constants
 \be\label{coeff0}
 c_1=E,~~c_2=L
 \ee
in terms of the conserved quantities $E$ and $L$. Finally, by
letting $ds^2=-kd\tau^2$ in Eq. (\ref{metric}) and making use of
Eqs. (\ref{v0s})--(\ref{coeff0}), one can obtain
 \be\label{rvel}
 v^1=\frac{dr}{d\tau}
    =\pm\left[E^2-\left(k+\frac{L^2}{r^2}\right)\left(1-\frac{2m}{r}\right)\right]^{1/2},
 \ee
where the $-(+)$ sign is for inward (outward) motion and $k=1(0)$
is for a timelike (nulllike) geodesic.


Now, let us find a freely falling acceleration and corresponding
temperature in the (5+1)-dimensional embedded flat spacetime,
which is described by Eq. (\ref{gems-sch}). For simplicity, we
assume that an observer is moving along a timelike geodesic with
zero angular momentum. Thus, we choose $k=1$ and $L=0$ in the
geodesic equations, which read as follows
 \bea
 \frac{dt}{d\tau}&=&\frac{E}{f(r)}, \\
 \frac{dr}{d\tau}&=&-\left[E^2-f(r)\right]^{1/2}.
 \eea

(a) A drip frame: for an observer who is freely falling from rest
$r=r_{0}$ at $\tau=0$, one has $E=\pm\sqrt{f(r_0)}$. Since this
case was already obtained, let us briefly summarize it
\cite{Brynjolfsson:2008uc,Kim:2009ha,Kim:2013wpa,Kim:2015wwa,hong2015}.
The equations of motion are reduced to
 \bea\label{eomr0}
 \frac{dt}{d\tau}&=&\frac{\sqrt{f(r_0)}}{f(r)},\nonumber\\
 \frac{dr}{d\tau}&=&-[f(r_0)-f(r)]^{1/2},
 \eea
where we have chosen the $(+)$ sign in $E$. Then, making use of
the embedding coordinates in Eq. (\ref{gems-sch}) and the geodesic
equations in Eq. (\ref{eomr0}), one can explicitly find a freely
falling acceleration $\bar{a}_{6}$ as
 \be\label{a6sch0}
 \bar{a}^2_{6}=\sum_{I=0}^{5}\left.\eta_{IJ}\frac{dz^I}{d\tau}\frac{dz^J}{d\tau}\right|_{r=r_0}
              =\frac{r^3+r_Hr^2+r^2_Hr+r^3_H}{4r^2_Hr^3}.
 \ee
This gives us the freely falling temperature measured by the
freely falling observer in the drip frame as
 \be \label{tffar-schw}
 T_{\rm DF}=\frac{\bar{a}_6}{2\pi}
           =\frac{1}{4\pi r_H}\sqrt{1+\frac{r_H}{r}+\frac{r^2_H}{r^2}+\frac{r^3_H}{r^3}}.
 \ee
Note that $r_0$ is replaced with $r$ in Eq. (\ref{a6sch0}). As
$r\rightarrow\infty$, the freely falling temperature $T_{\rm DF}$
is reduced to the Hawking temperature (\ref{HawkingT}). Note also
that at the event horizon the freely falling temperature
(\ref{tffar-schw}) is finite, while the fiducial temperature
(\ref{fidT}) diverges \cite{Brynjolfsson:2008uc,Kim:2009ha}.

(b) A rain frame: for an observer who is freely falling from rest
at the asymptotic infinity, one has $E=1$.  Then, the equations of
motion are reduced to
 \bea\label{eomr0s-rf}
 \frac{dt}{d\tau}&=&\frac{r}{r-2m},\nonumber\\
 \frac{dr}{d\tau}&=&-\sqrt{\frac{2m}{r}}.
 \eea
Then, in this case, making use of the embedding coordinates in Eq.
(\ref{gems-sch}) and the geodesic equations in Eq. (\ref{eomr0}),
one can obtain a freely falling acceleration $\bar{a}_{6}$
 \be
 \bar{a}^2_{6}=\sum_{I=0}^{5}\left.\eta_{IJ}\frac{dz^I}{d\tau}\frac{dz^J}{d\tau}\right|_{r=\infty}
               =\frac{1}{4r^2_H},
 \ee
and thus the freely falling temperature as
 \be
 T_{\rm RF}=\frac{\bar{a}_6}{2\pi}=\frac{1}{4\pi r_H}.
 \ee
Note that the acceleration $\bar{a}_6$ is only defined at spatial
infinity. Thus, $T_{\rm RF}$ is the freely falling temperature
$T_{\rm DF}$ in Eq. (\ref{tffar-schw}) with $r\rightarrow\infty$.
Therefore, one can see that the Hawking temperature in a curved
spacetime is equal to the Unruh temperature in a
higher-dimensional flat spacetime having a freely falling
acceleration starting from rest at the asymptotic infinity.

(c) A hail frame: finally, if an observer starts to fall with an
inward non-zero velocity at the asymptotic infinity as
 \be\label{vel-hail}
 \frac{dr}{dt}=-v_\infty
 \ee
where the $(-)$ sign is for the inward direction, one has
$E=(1-v^2_\infty)^{-1/2}$.  Then, the equations of motion are
reduced to
 \bea\label{eomr0s-hf}
 \frac{dt}{d\tau}&=&\frac{1}{\sqrt{1-v^2_\infty}\left(1-\frac{2m}{r}\right)},\nonumber\\
 \frac{dr}{d\tau}&=&-\left(\frac{v^2_\infty}{1-v^2_\infty}+\frac{2m}{r}\right)^{1/2}.
 \eea
It seems to appropriate to comment that the equations of motion
(\ref{eomr0s-hf}) give
 \be
 \left.\frac{dr}{dt}\right|_{r=r_H}=-1
 \ee
at the event horizon, which shows that the velocity of the
observer does not exceed the speed of light \cite{Tayler:2000},
independent of how fast the observer is hurled at the asymptotic
infinity given by (\ref{vel-hail}).

Now, by making use of the embedding coordinates in Eq.
(\ref{gems-sch}) and the geodesic equations in Eq.
(\ref{eomr0s-hf}), we can obtain an acceleration $\bar{a}_{6}$ as
 \be
 \bar{a}^2_{6}=\sum_{I=0}^{5}\left.\eta_{IJ}\frac{dz^I}{d\tau}\frac{dz^J}{d\tau}\right|_{r=\infty}
               = \frac{1}{4r^2_H(1-v^2_\infty)^2}.
 \ee
Thus, the freely falling temperature in the hail frame is obtained
as
 \be\label{hailT}
 T_{\rm HF}=\frac{\bar{a}_6}{2\pi}=\frac{1}{4\pi r_H(1-v^2_\infty)}.
 \ee
Note that when $v_\infty=0$, this temperature in the hail frame is
reduced to the freely falling one in the rain frame as expected.
The freely falling temperature in the hail frame can also be
rewritten as
 \be
T_{\rm HF}=\frac{E^2}{4\pi r_H}
 \ee
in terms of the energy per unit mass. It is appropriate to comment
that even though we may have the freely falling temperature in the
hail frame, it would not be actually well-defined in the hail
frame since the observer is not at rest even momentarily.

\subsection{GEMS of the Schwarzschild black hole in the PG spacetime}

In this subsection, we newly find GEMS embeddings of the
(3+1)-dimensional Schwarzschild black hole in the PG spacetime,
comparing with the previous results of the Schwarzschild black
hole in the spherically symmetric spacetime. In the PG spacetime,
GEMS embedding is obtained by redefining the PG time
\cite{Martel:2000rn} as
 \be
 t=t_S-F_S(r),
 \ee
where $t_S$ is the Schwarzschild time written in the metric
(\ref{metric}) and
 \be
 F_S(r)=-\int\sqrt{\frac{2m}{r}}\left(1-\frac{2m}{r}\right)^{-1}dr.
 \ee
This integral can be easily performed and one has
 \be\label{fs}
 F_S(r)=-4m\left(y_s-\frac{1}{2}\ln\frac{y_s+1}{y_s-1}\right)
 \ee
with the definition
 \be
 y_s\equiv (2m/r)^{-1/2}.
 \ee
Thus, the Schwarzschild black hole in the PG spacetime is
described by the metric
 \be\label{pgmetric}
  ds^2=-\left(1-\frac{2m}{r}\right)dt^2+2\sqrt{\frac{2m}{r}}dtdr+dr^2+r^2(d\theta^2+\sin^2\theta
  d\phi^2),
 \ee
which has an off-diagonal term. This PG spacetime is stationary,
but not static. Note also that there is no coordinate singularity
at $r_H=2m$, which is the Schwarzschild radius. Unlike the
Schwarzschild coordinate, it is possible to define an effective
vacuum state of a quantum field with this well-behaved coordinate
system at the horizon by requiring that it annihilates negative
frequency modes with respect to the PG time $t$. Moreover,
positive energy particles can tunnel out through a barrier set by
the energies of outgoing particles themselves
\cite{Kraus:1994fh,Parikh:1999mf,Parikh:2004ih}. Observers in this
PG spacetime, who are freely falling from rest at infinity and see
nothing abnormal at the horizon, carry their own proper times.
Each spatial slice of the metric with $dt=0$ corresponds to the
flat metric in the spherical coordinates. Thus, the spacetime
curvature information is contained in the off-diagonal component
of the metric (\ref{pgmetric}), which structure we will study in
the following GEMS method in detail.

Inspired by the previous GEMS scheme of the Schwarzschild
spacetime, the (3+1)-dimensional Schwarzschild black hole in the
PG spacetime can be embedded into a (5+1)-dimensional flat one
described as
 \be
 ds^{2}= \eta_{IJ}dz^Idz^J,~{\rm with}~\eta_{IJ}={\rm diag}(-1,1,1,1,1,1),
 \ee
with
 \bea \label{gemsgp}
 z^{0}&=&k_{H}^{-1}f^{1/2}(r)\sinh k_{H}(t+F_S(r)), \nonumber \\
 z^{1}&=&k_{H}^{-1}f^{1/2}(r)\cosh k_{H}(t+F_S(r)), \nonumber \\
 z^{2}&=&r\sin\theta\cos\phi, \nonumber \\
 z^{3}&=&r\sin\theta\sin\phi, \nonumber \\
 z^{4}&=&r\cos\theta, \nonumber\\
 z^{5}&=&\int dr
 \left(\frac{r_H(r^2+rr_H+r^2_H)}{r^3}\right)^{1/2},
 \eea
where $f(r)$ is denoted in Eq. (\ref{fr}) and $k_H=1/4m$ can be
calculated directly from the metric (\ref{pgmetric}) with the
definition of (\ref{sgravity}). Note that compared with the GEMS
embeddings in Eq. (\ref{gems-sch}), the coordinates $z^0$ and
$z^1$ are transformed to include the PG time $t$, which partly
lead to the off-diagonal term in the PG metric (\ref{pgmetric}).
On the other hand, the coordinates $z^2$, $z^3$, and $z^4$
constitute the spatial hypersurface, which is the same with the
constant time spacial slice of the metric (\ref{pgmetric}).

Now, in static detectors ($r$, $\theta$, $\phi=$constant)
described by a fixed point in the ($z^{2}$, $z^{3}$, $z^{4}$,
$z^{5}$) plane, a uniformly accelerated observer in the
(5+1)-dimensional flat spacetime, follows a hyperbolic trajectory
in ($z^{0}$,$z^{1}$) described by a proper acceleration $a_{6}$ as
follows
 \be\label{acc-msch}
 {a}_{6}^{-2}=(z^1)^2-(z^0)^2
  =\frac{f(r)}{k^2_H}.
 \ee
Thus, we arrive at the Unruh temperature for the uniformly
accelerated observer in the (5+1)-dimensional flat spacetime
 \be\label{ut-msch}
 T_U=\frac{a_6}{2\pi}=\frac{k_H}{2\pi\sqrt{f(r)}}.
 \ee
It is also appropriate to comment that the static detectors having
the same condition ($r,~\theta,~\phi={\rm constant}$) in this PG
spacetime following timelike Killing vector $\xi=\partial_t$ have
the same constant 4-acceleration (\ref{a4sch}) and thus satisfy
the same relation (\ref{a6a4}).

Now, let us consider the geodesic equation (\ref{gdeq}) in the
Schwarzschild black hole in the PG spacetime. From the metric
(\ref{pgmetric}), one can obtain the geodesic equations explicitly
as
 \bea
 &&\frac{dv^0}{d\tau}+\frac{m}{r^2}\sqrt{\frac{2m}{r}}(v^0)^2+\frac{2m}{r^2}v^0v^1
    +\frac{1}{r}\sqrt{\frac{m}{2r}}(v^1)^2
    -\sqrt{2mr}[(v^2)^2+\sin^2\theta(v^3)^2]=0, \\
 &&\frac{dv^1}{d\tau}+\frac{m}{r^3}(r-2m)(v^0)^2
       -\frac{2m}{r^2}\sqrt{\frac{2m}{r}}v^0v^1
       -\frac{m}{r^2}(v^1)^2-(r-2m)\left[(v^2)^2+\sin^2\theta(v^3)^2\right]=0,     \\
 &&\frac{dv^2}{d\tau}+\frac{2}{r}v^1v^2-\sin\theta\cos\theta(v^3)^2=0, \\
 &&\frac{dv^3}{d\tau}+\frac{2}{r}v^1v^3+2\cot\theta v^2 v^3=0.
 \eea
As before, for simplicity, one can consider the geodesics on the
equatorial plane $\theta=\pi/2$. Then, the geodesic equations are
reduced to
 \bea
 &&\frac{dv^0}{d\tau}+\frac{m}{r^2}\sqrt{\frac{2m}{r}}(v^0)^2+\frac{2m}{r^2}v^0v^1
  +\frac{1}{r}\sqrt{\frac{m}{2r}}(v^1)^2
    -\sqrt{2mr}(v^3)^2=0, \label{ge0pg}\\
 &&\frac{dv^1}{d\tau}+\frac{m}{r^3}(r-2m)(v^0)^2
       -\frac{2m}{r^2}\sqrt{\frac{2m}{r}}v^0v^1
        -\frac{m}{r^2}(v^1)^2
       -(r-2m)(v^3)^2=0,     \\
 &&\frac{dv^3}{d\tau}+\frac{2}{r}v^1v^3=0.\label{ge2pg}
 \eea
First of all, one can easily integrate (\ref{ge2pg}) as
 \bea
 v^3&=&\frac{d\phi}{d\tau}=\frac{c_3}{r^2}, \label{v2pg}
 \eea
where $c_3$ is an integration constant. Note that as before, for
the Killing vectors $\xi^\mu=(1,0,0,0)$ and $\psi^\mu=(0,0,0,1)$,
one can have conserved quantities of $E$ and $L$ given by
 \bea
 E &=& -g_{\mu\nu}\xi^\mu v^\nu =\left(1-\frac{2m}{r}\right) v^0
                                 -\sqrt{\frac{2m}{r}}v^1,\label{K1pg}\\
 L &=& g_{\mu\nu}\psi^\mu v^\nu =r^2 v^3.\label{K2pg}
 \eea
Comparing (\ref{K2pg}) with Eq. (\ref{v2pg}), one can fix the
integration constant as
 \be\label{coeff0pg}
 c_3=L
 \ee
in terms of the conserved quantity $L$. Also, by letting
$ds^2=-kd\tau^2$ in Eq. (\ref{pgmetric}) and making use of Eqs.
(\ref{v2pg}) and (\ref{K1pg}), one can obtain
 \be\label{rvelpg}
 v^1=\frac{dr}{d\tau}
    =\pm\left[E^2-\left(k+\frac{L^2}{r^2}\right)\left(1-\frac{2m}{r}\right)\right]^{1/2},
 \ee
where the $-(+)$ sign is for inward (outward) motion and $k=1(0)$
is for a timelike (nulllike) geodesic. Finally, making use of Eq.
(\ref{rvelpg}), one can obtain
 \be
 v^0=\frac{dt}{d\tau}
    =\frac{E-\sqrt{\frac{2m}{r}}\left[E^2-\left(k+\frac{L^2}{r^2}\right)
         \left(1-\frac{2m}{r}\right)\right]^{1/2}}{1-\frac{2m}{r}}.
 \ee
Here, we have used the $(-)$ sign for $v^1$ describing the inward
motion for later convenience.

Now, let us find a freely falling acceleration and corresponding
temperature in the (5+1)-dimensional embedded flat spacetime. As
in the previous subsection, for simplicity, we assume that an
observer is moving along a timelike geodesic with zero angular
momentum. Thus, we choose $k=1$ and $L=0$ in the geodesic
equations, which read as follows
 \bea
 \frac{dt}{d\tau}&=&\frac{E-\sqrt{\frac{2m}{r}[E^2-f(r)]}}{f(r)}, \\
 \frac{dr}{d\tau}&=&-\left[E^2-f(r)\right]^{1/2}.
 \eea
(a) A drip frame: for an observer who is freely falling from rest
$r=r_{0}$ at
$\tau=0$~\cite{Brynjolfsson:2008uc,Kim:2009ha,Kim:2013wpa,Kim:2015wwa,hong2015},
one has $E=\pm\sqrt{f(r_0)}$. Then, the equations of motion are
reduced to
 \bea\label{eomr0pg}
 \frac{dt}{d\tau}&=&\frac{\sqrt{f(r_0)}-\sqrt{\frac{2m}{r}\left[f(r_0)-f(r)\right]}}{f(r)},\nonumber\\
 \frac{dr}{d\tau}&=&-[f(r_0)-f(r)]^{1/2}.
 \eea
Note that we have chosen the $(+)$ sign in $E$.

Now, by exploiting the embedding coordinates in Eq. (\ref{gemsgp})
and the equations of motion in Eq. (\ref{eomr0pg}), one can easily
obtain a freely falling acceleration $\bar{a}_{6}$ in the
(5+1)-dimensional GEMS embedded spacetime as
 \be\label{a6sch}
 \bar{a}^2_{6}=\sum_{I=0}^{5}\left.\eta_{IJ}\frac{dz^I}{d\tau}\frac{dz^J}{d\tau}\right|_{r=r_0}
              =\frac{r^3+r_Hr^2+r^2_Hr+r^3_H}{4r^2_Hr^3},
 \ee
which is exactly the same with the freely falling acceleration
(\ref{a6sch0}) as expected. Thus, the freely falling temperature
at rest measured by the freely falling observer can be written as
 \be \label{tffar-sch}
 T_{\rm DF}=\frac{\bar{a}_6}{2\pi}
  = \frac{1}{4\pi r_H}\sqrt{1+\frac{r_H}{r}+\frac{r^2_H}{r^2}+\frac{r^3_H}{r^3}}.
 \ee

\begin{figure*}[t!]
   \centering
   \includegraphics{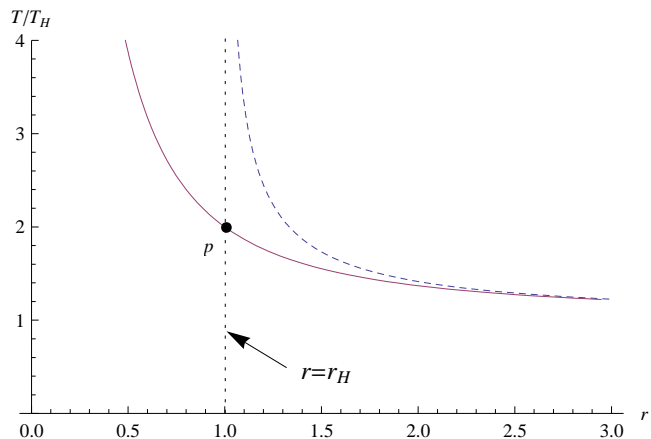}
\caption{Freely falling temperature in a drip frame (solid line)
and fiducial temperature (dashed line) in the PG coordinates in
the unit of the Hawking temperature with $r_H=1$. For fiducial
observers, freely falling temperature is seen to diverge at the
event horizon. However, for freely falling observers in a drip
frame, freely falling temperature is well-defined at the horizon
and increased to the $r=0$ singularity.}
 \label{fig1}
\end{figure*}

In Fig. \ref{fig1}, we have drawn the freely falling and fiducial
temperatures in the unit of the Hawking temperature. For fiducial
observers at rest from a distance in the PG coordinates, freely
falling temperature is seen to diverge at the event horizon.
However, for freely falling observers in the PG coordinates,
freely falling temperature is well-defined at the horizon and
moreover is continuously increased to the $r=0$ singularity by
passing through the horizon. It is appropriate to comment that if
the observers are at rest from a distance in the usual
Schwarzschild coordinates as in the previous subsection, this
freely falling temperature would be ended at finite temperature at
the event horizon which is denoted by the point $p$ in Fig.
\ref{fig1}.

(b) A rain frame: for an observer who is freely falling from rest
at the asymptotic infinity, one has $E=1$. Then, the equations of
motion are reduced to
 \bea\label{eomr0pg-rf}
 \frac{dt}{d\tau}&=&1,\nonumber\\
 \frac{dr}{d\tau}&=&-\sqrt{\frac{2m}{r}}.
 \eea
Thus, one can obtain a freely falling acceleration $\bar{a}_{6}$
as
 \be
 \bar{a}^2_{6} =\sum_{I=0}^{5}\left.\eta_{IJ}\frac{dz^I}{d\tau}\frac{dz^J}{d\tau}\right|_{r=\infty}
              =\frac{1}{4r^2_H},
 \ee
and the freely falling temperature at rest at the asymptotic
infinity as
 \be
 T_{\rm RF}=\frac{1}{4\pi r_H},
 \ee
which is exactly the same with the Hawking temperature
(\ref{HawkingT}) as before.

It is appropriate to comment that four velocity in the rain frame
can be obtained from the drip frame by taking limit of
$r_{0}\rightarrow \infty$, which is given by
 \be \label{uasch}
 v^{a}=\left(1,-\sqrt{\frac{2m}{r}},0,0\right),
 \ee
which shows that the observer's time is given by the proper time.
Then, the velocity seen by the observer in the asymptotic infinity
becomes
 \be
 \frac{dr}{dt}=-\sqrt{\frac{2m}{r}},
 \ee
which is the escape velocity. Thus, one can see that the observer
is freely falling with the flat space slicing which is flowing
radially inwards at the escape velocity. Note that at the event
horizon $r_H=2m$, it becomes the speed of light.

(c) A hail frame: for an observer who has thrown with a velocity
$v_\infty$ at the asymptotic infinity as
 \be
 \frac{dr}{dt}=-v_\infty,
 \ee
one has $E=(1-v^2_\infty)^{-1/2}$. Then, the equations of motion
are reduced to
 \bea\label{eomr0pg-hf}
 \frac{dt}{d\tau}&=&\frac{(1-v^2_\infty)^{-1/2}
           -\sqrt{\frac{2m}{r}
                \left(\frac{v^2_\infty}{1-v^2_\infty}+\frac{2m}{r}\right)}}
           {1-\frac{2m}{r}},\nonumber\\
 \frac{dr}{d\tau}&=&-\left(\frac{v^2_\infty}{1-v^2_\infty}+\frac{2m}{r}\right)^{1/2}.
 \eea

In this case, one can obtain an acceleration $\bar{a}_{6}$ by
making use of the embedding coordinates in Eq. (\ref{gemsgp}) as
 \be
 \bar{a}^2_{6} =\sum_{I=0}^{5}\left.\eta_{IJ}\frac{dz^I}{d\tau}\frac{dz^J}{d\tau}\right|_{r=\infty}
              =\frac{1}{4r^2_H(1-v^2_\infty)^2}.
 \ee
Thus, the freely falling-like temperature in the hail frame is
obtained as
 \be
 T_{\rm HF}=\frac{1}{4\pi r_H(1-v^2_\infty)}.
 \ee
Note that for the case of $v_\infty=0$, $T_{\rm HF}$ reduces to
$T_{\rm RF}$ at infinity and as the throwing velocity $v_\infty$
is increased, $T_{\rm HF}$ is also increased. As before, the
freely falling temperature in the hail frame can also be rewritten
as
 \be
  T_{\rm HF}=\frac{E^2}{4\pi r_H}
 \ee
in terms of the energy per unit mass.

\section{GEMS of the RN black hole in the PG spacetime}
\setcounter{equation}{0}
\renewcommand{\theequation}{\arabic{section}.\arabic{equation}}

Let us consider the RN black hole in the spherically symmetric
spacetime
 \be\label{metric-rn}
 ds^2=-f(r)dt^2_{RN}+f^{-1}(r)dr^2+r^2(d\theta^2+\sin^2\theta d\phi^2)
 \ee
with
 \be\label{frn0}
 f(r)=1-\frac{2m}{r}+\frac{q^2}{r^2}.
 \ee
It was well-known that this RN spacetime can be embedded into the
(5+2)-dimensional flat spacetime
\cite{Deser:1998xb,Brynjolfsson:2008uc,Kim:2000ct,Paston:2013mfa,Kim:2009ha,Banerjee:2010ma}.
So, we do not repeat it. Instead, in the following, we will
concentrate on GEMS embedding of the RN black hole in the PG
spacetime.

Now, by defining a new time in the PG coordinates as
 \be
 t=t_{RN}-F_{RN}(r),
 \ee
where $t_{RN}$ is the time in the usual RN spacetime and
 \be\label{FRN}
 F_{RN}(r)=-\int\sqrt{\frac{2m}{r}-\frac{q^2}{r^2}}
            \left(1-\frac{2m}{r}+\frac{q^2}{r^2}\right)^{-1}dr,
 \ee
one can obtain the RN black hole in the PG spacetime as
 \be\label{pgmetric-rn}
 ds^2=-\left(1-\frac{2m}{r}+\frac{q^2}{r^2}\right)dt^2+2\sqrt{\frac{2m}{r}-\frac{q^2}{r^2}}dtdr
       +dr^2+r^2(d\theta^2+\sin^2\theta d\phi^2),
 \ee
which also has an off-diagonal term as before.

First of all, it seems appropriate to comment on the integral
(\ref{FRN}). By redefining
 \be\label{RNyq}
 y_q\equiv\left(\frac{2m}{r}-\frac{q^2}{r^2}\right)^{-1/2},
 \ee
let us rewrite $F_{RN}(r)$ as
 \be
 F_{RN}(r)=-4m\int dy_q \frac{y^2_q}{y^2_q-1}(1-\alpha)^{-1},
 \ee
with $\alpha=(qy_q/r)^4$. Then, through tedious calculations, we
have newly obtained the exact solution of the nontrivial form
 \bea
 F_{RN}(r)&=&F^q_S(r)
          - 8m\sum^\infty_{k=1}\frac{(k-1)(2k-3)!}{k!(k-2)!}
           \left(\frac{q}{2m}\right)^{2k}
           \left(\sum^{k-2}_{n=0}\frac{y^{-(2n+1)}_q}{2n+1}-\frac{1}{2}\ln\frac{y_q+1}{y_q-1}\right),
 \eea
where
 \be\label{fsrn}
 F^q_S(r)=-4m\left(y_q-\frac{1}{2}\ln\frac{y_q+1}{y_q-1}\right).
 \ee
Note that in $y_q$ one should require the condition of $2mr>q^2$.
Otherwise, there is an obstruction to implement spatially flat PG
hypersurfaces \cite{Lin:2008xg}. In the limit of $q\rightarrow 0$,
$F_{RN}(r)$ becomes $F_S(r)$ in Eq. (\ref{fs}) as expected.

Now, exploiting the GEMS approach, we can embed the
(3+1)-dimensional RN black hole in the PG spacetime into one in a
(5+2)-dimensional flat spacetime as
 \be
 ds^{2}= \eta_{IJ}dz^Idz^J,~{\rm with}~\eta_{IJ}={\rm diag}(-1,1,1,1,1,1,-1),
 \ee
where the embedding coordinates are explicitly given by
 \bea\label{gems-rngp}
 z^{0}&=&k_{H}^{-1}f^{1/2}(r)\sinh k_{H}(t+F_{RN}(r)), \nonumber \\
 z^{1}&=&k_{H}^{-1}f^{1/2}(r)\cosh k_{H}(t+F_{RN}(r)), \nonumber \\
 z^{2}&=&r\sin\theta\cos\phi, \nonumber \\
 z^{3}&=&r\sin\theta\sin\phi, \nonumber \\
 z^{4}&=&r\cos\theta, \nonumber \\
 z^{5}&=&\int dr \left(\frac{r^2(r_++r_-)+r^2_+(r_++r_-)}{r^2(r-r_-)}\right)^{1/2},\nonumber\\
 z^{6}&=&\int dr \left(\frac{4r^5_+r_-}{r^4(r_+-r_-)^2}\right)^{1/2}.
 \eea
Here, $k_H$ is given by
 \be
 k_H =\frac{r_+-r_-}{2r^2_+}
 \ee
with $r_\pm=m\pm\sqrt{m^2-q^2}$. Again, compared with the previous
GEMS embeddings of the RN black hole in the spherically symmetric
spacetime
\cite{Deser:1998xb,Brynjolfsson:2008uc,Kim:2000ct,Paston:2013mfa,
Kim:2009ha,Banerjee:2010ma}, the coordinates $z^0$ and $z^1$
include the new PG time $t$, which partly lead to the off-diagonal
term in the PG metric (\ref{pgmetric-rn}). And the coordinates
$z^2$, $z^3$, and $z^4$ constitute the flat spatial hypersurfaces
as far as we keep the condition of $2mr>q^2$.

In static detectors ($r,~\theta,~\phi={\rm constant}$) described
by a fixed point in the ($z^2,~z^3,~z^4,~z^5,~z^6$) plane, a
uniformly accelerated observer in the (5+2)-dimensional flat
spacetime, follows a hyperbolic trajectory in ($z^0,~z^1$)
described by
 \be\label{a7rn0}
 a^{-2}_7=(z^1)^2-(z^0)^2= \frac{4r^4_+(r-r_+)(r-r_-)}{r^2(r_+-r_-)^2}.
 \ee
Thus, we arrive at the Unruh temperature for a uniformly
accelerated observer in the (5+2)-dimensional flat spacetime
 \be\label{ut-rn}
 T_U=\frac{a_7}{2\pi}=\frac{r(r_+-r_-)}{4\pi r^2_+(r-r_+)^{1/2}(r-r_-)^{1/2}}.
 \ee

It is also appropriate to comment that the static detectors having
the same condition ($r,~\theta,~\phi={\rm constant}$) in the
original (3+1)-dimensional spacetime following the timelike
Killing vector $\xi=\partial_t$ have the constant 4-acceleration
as
 \be\label{a4rn}
 a_4=\frac{(r_++r_-)r-2r_+r_-}{2r^2[(r-r_+)(r-r_-)]^{1/2}}.
 \ee
Comparing this with the acceleration (\ref{a7rn0}) in the
(5+2)-dimensional embedded spacetime, one can have
 \be\label{a7a4}
 a_7=\frac{r^3(r_+-r_-)}{r^2_+[(r_++r_-)r-2r_+r_-]}a_4.
 \ee
In the Schwarzschild limit of $r_+=r_H$ and $r_-=0$, the
acceleration $a_7$ becomes $a_6$, while $a_4$ becomes
(\ref{a4sch}) so that the relation (\ref{a6a4}) is recovered as
expected.

Now, as in the previous section, let us consider the geodesic
equation (\ref{gdeq}) in the RN black hole in the PG spacetime.
First of all, non-vanishing independent components of the
Christoffel symbols are given by
 \bea
 \Gamma^0_{00}&=& -\Gamma^1_{01}=\frac{1}{r}\left(\frac{m}{r}-\frac{q^2}{r^2}\right)\left(\frac{2m}{r}-\frac{q^2}{r^2}\right)^{1/2},
               ~~~~~\Gamma^0_{01}=-\Gamma^1_{11}=\frac{1}{r}\left(\frac{m}{r}-\frac{q^2}{r^2}\right),\nonumber\\
 \Gamma^0_{11}&=& \frac{1}{r}\left(\frac{m}{r}-\frac{q^2}{r^2}\right)\left(\frac{2m}{r}-\frac{q^2}{r^2}\right)^{-1/2},
               ~~~~~~~~~~~\Gamma^0_{22}=-r\left(\frac{2m}{r}-\frac{q^2}{r^2}\right)^{1/2},\nonumber\\
 \Gamma^0_{33}&=&-r\left(\frac{2m}{r}-\frac{q^2}{r^2}\right)^{1/2}\sin^2\theta,
               ~~~~~~~~~~~~~~~~~~~~\Gamma^1_{00}=\frac{1}{r}\left(\frac{m}{r}-\frac{q^2}{r^2}\right)\left(1-\frac{2m}{r}+\frac{q^2}{r^2}\right),\nonumber\\
 \Gamma^1_{22}&=&-r\left(1-\frac{2m}{r}+\frac{q^2}{r^2}\right),
               ~~~~~~~~~~~~~~~~~~~~~~~~~~~\Gamma^1_{33}=-r\left(1-\frac{2m}{r}+\frac{q^2}{r^2}\right)\sin^2\theta, \nonumber\\
 \Gamma^2_{12}&=&\Gamma^3_{13}=\frac{1}{r},
               ~~~~~~~~~~~~~~~~~~~~~~~~~~~~~~~~~~~~~~~~~\Gamma^2_{33}=-\sin\theta\cos\theta,\nonumber\\
 \Gamma^3_{23}&=&\cot\theta.
 \eea
Then, from the metric (\ref{pgmetric-rn}), one can obtain the
geodesic equations explicitly as
 \bea
 &&\frac{dv^0}{d\tau}
    +\frac{\sqrt{2mr-q^2}(mr-q^2)}{r^4}(v^0)^2
    +\frac{2(mr-q^2)}{r^3}v^0v^1
    +\frac{mr-q^2}{r^2\sqrt{2mr-q^2}}(v^1)^2 \nonumber\\
  &&  -\sqrt{2mr-q^2}[(v^2)^2+\sin^2\theta(v^3)^2]=0, \\
 &&\frac{dv^1}{d\tau}
       +\frac{[(r-2m)r+q^2](mr-q^2)}{r^5}(v^0)^2
       -\frac{2\sqrt{2mr-q^2}(mr-q^2)}{r^4}v^0v^1
         -\frac{mr-q^2}{r^3}(v^1)^2
      \nonumber  \\
  &&
        -\frac{r^2-2mr+q^2}{r}\left[(v^2)^2+\sin^2\theta(v^3)^2\right]=0,     \\
 &&\frac{dv^2}{d\tau}+\frac{2}{r}v^1v^2-\sin\theta\cos\theta(v^3)^2=0, \\
 &&\frac{dv^3}{d\tau}+\frac{2}{r}v^1v^3+2\cot\theta v^2 v^3=0.
 \eea
As before, for simplicity, one can consider the geodesics on the
equatorial plane $\theta=\pi/2$. Then, the geodesic equations are
reduced to
 \bea
 &&\frac{dv^0}{d\tau}
    +\frac{\sqrt{2mr-q^2}(mr-q^2)}{r^4}(v^0)^2
    +\frac{2(mr-q^2)}{r^3}v^0v^1
    +\frac{mr-q^2}{r^2\sqrt{2mr-q^2}}(v^1)^2\nonumber \\
  &&  -\sqrt{2mr-q^2}(v^3)^2=0, \label{ge0rn}\\
 &&\frac{dv^1}{d\tau}
        +\frac{[(r-2m)r+q^2](mr-q^2)}{r^5}(v^0)^2
       -\frac{2\sqrt{2mr-q^2}(mr-q^2)}{r^4}v^0v^1
       -\frac{mr-q^2}{r^3}(v^1)^2
       \nonumber\\
  &&
        -\frac{r^2-2mr+q^2}{r}(v^3)^2=0,     \\
 &&\frac{dv^3}{d\tau}+\frac{2}{r}v^1v^3=0.\label{ge2rn}
 \eea
Firstly, one can easily integrate (\ref{ge2rn}) as
 \bea
 v^3&=&\frac{d\phi}{d\tau}=\frac{c_4}{r^2}, \label{v2rn}
 \eea
where $c_4$ is an integration constant. Now, for the Killing
vectors $\xi^\mu=(1,0,0,0)$ and $\psi^\mu=(0,0,0,1)$, one can have
conserved quantities of $E$ and $L$ given by
 \bea
 E &=& -g_{\mu\nu}\xi^\mu v^\nu =\left(1-\frac{2m}{r}+\frac{q^2}{r^2}\right) v^0
                                 -\sqrt{\frac{2m}{r}-\frac{q^2}{r^2}}v^1,\label{K1rn}\\
 L &=& g_{\mu\nu}\psi^\mu v^\nu =r^2 v^3.\label{K2rn}
 \eea
Comparing (\ref{K2rn}) with Eq. (\ref{v2rn}), one can fix the
integration constant as
 \be\label{coeff0rn}
 c_4=L
 \ee
in terms of the conserved quantity $L$. Also, by letting
$ds^2=-kd\tau^2$ in Eq. (\ref{metric-rn}) and making use of Eqs.
(\ref{v2rn}) and (\ref{K1rn}), one can obtain
 \be\label{rvelrn}
 v^1=\frac{dr}{d\tau}
    =\pm\left[E^2-\left(k+\frac{L^2}{r^2}\right)\left(1-\frac{2m}{r}+\frac{q^2}{r^2}\right)\right]^{1/2},
 \ee
where the $-(+)$ sign is for inward (outward) motion and $k=1(0)$
is for a timelike (nulllike) geodesic. Finally, making use of
(\ref{rvelrn}), one can obtain
 \be
 v^0=\frac{dt}{d\tau}
    =\frac{E-\sqrt{\frac{2m}{r}-\frac{q^2}{r^2}}
         \left[E^2-\left(k+\frac{L^2}{r^2}\right)
         \left(1-\frac{2m}{r}+\frac{q^2}{r^2}\right)\right]^{1/2}}
     {1-\frac{2m}{r}+\frac{q^2}{r^2}}.
 \ee
As before, we have here used the $(-)$ sign for $v^1$ describing
the inward motion for later convenience.

Now, let us find a freely falling acceleration and corresponding
temperature in the (5+2)-dimensional embedded flat spacetime. As
in the previous section, for simplicity, we assume that an
observer is moving along a timelike geodesic with zero angular
momentum. Thus, we choose $k=1$ and $L=0$ in the geodesic
equations, which read as follows
 \bea
 \frac{dt}{d\tau}&=&\frac{E-\sqrt{(\frac{2m}{r}-\frac{q^2}{r^2})[E^2-f(r)]}}{f(r)}, \\
 \frac{dr}{d\tau}&=&-\left[E^2-f(r)\right]^{1/2}.
 \eea

(a) A drip frame: for an observer who is freely falling from rest
$r=r_{0}$ at
$\tau=0$~\cite{Brynjolfsson:2008uc,Kim:2009ha,Kim:2013wpa,Kim:2015wwa,hong2015},
one has $E=\pm\sqrt{f(r_0)}$. Then, the equations of motion are
reduced to
 \bea\label{eomr0rn-df}
 \frac{dt}{d\tau}&=&\frac{\sqrt{f(r_0)}-\sqrt{\left(\frac{2m}{r}-\frac{q^2}{r^2}\right)\left[f(r_0)-f(r)\right]}}{f(r)},\nonumber\\
 \frac{dr}{d\tau}&=&-[f(r_0)-f(r)]^{1/2}.
 \eea
Note that we have chosen the $(+)$ sign in $E$.

Now, by exploiting the embedding coordinates in Eq.
(\ref{gems-rngp}) and the equations of motion in Eq.
(\ref{eomr0rn-df}), one can easily obtain a freely falling
acceleration $\bar{a}_{7}$ in the (5+2)-dimensional GEMS embedded
spacetime as
 \be\label{a7rn}
 \bar{a}^2_{7}=\sum_{I=0}^{6}\left.\eta_{IJ}\frac{dz^I}{d\tau}\frac{dz^J}{d\tau}\right|_{r=r_0}
              =\frac{(r_+-r_-)^2(r^3+r_+r^2+r^2_+r+r^3_+)r^2-4r^5_+r_-(r-r_-)}{4(r-r_-)r^4_+r^4}.
 \ee
In the last expression, we have replaced $r_0$ with $r$. Thus, the
freely falling temperature at rest measured by the freely falling
observer can be written as
 \be \label{tffar-rn}
 T_{\rm DF}=\frac{\bar{a}_7}{2\pi}
           =\frac{(r_+-r_-)}{4\pi r^2_+}\sqrt{\frac{r^3+r_+r^2+r^2_+r+r^3_+}{{(r-r_-)r^2}}-\frac{4r^5_+r_-}{(r_+-r_-)^2r^4}}.
 \ee
In the limit of $r \rightarrow \infty$, the above $T_{\rm DF}$ is
reduced to the Hawking temperature
 \be
 T_{H}=\frac{r_{+}-r_{-}}{4\pi r_{+}^{2}}. \ee
For the case of $q=0$ ($r_+=r_H$, $r_-=0$), $ T_{\rm DF}$ becomes
the Schwarzschild one in Eq. (\ref{tffar-sch}). It also seems
appropriate to comment that when the observer approaches to the
event horizon, $T_{\rm DF}$ does not diverge but is finite as
 \be
 T_{\rm DF}=\frac{\sqrt{r_+(r_+-2r_-)}}{2\pi r^2_+},
 \ee
while the fiducial temperature (\ref{ut-rn}) diverges.

\begin{figure*}[t!]
   \centering
   \includegraphics{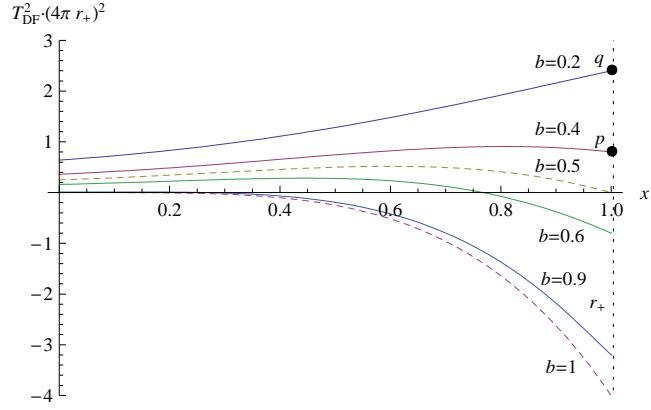}
\caption{Squared freely falling temperatures in a drip frame,
$T^2_{\rm DF}\cdot(4\pi r^2_+)^2$ along $x=r_+/r$ for
$b=0.2,~0.4,~0.5,~0.6,~0.9,{\rm and}~1$, where $b=r_-/r_+$. The
dotted line at $x=1$ is for the horizon $r_+$.}
 \label{fig2}
\end{figure*}

\begin{figure*}[t!]
   \centering
   \includegraphics{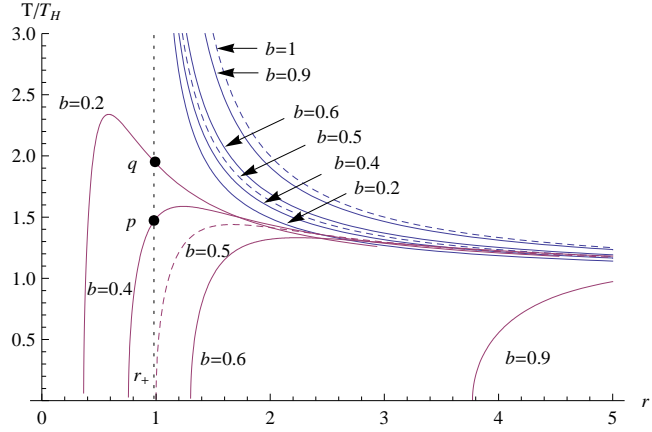}
\caption{Freely falling temperatures $T_{\rm DF}/T_H$ in the unit
of the Hawking temperature in a drip frame (lower curves) and
fiducial temperatures $T_{\rm FID}/T_H$  (upper curves) along $r$
for $b=0.2,~0.4,~0.5, ~0.6,~0.9,~{\rm and}~ 1$. The dotted line at
$r=1$ is for the horizon $r_+$. Note that there is no freely
falling temperature for $b=1$, which is imaginary for all $x$ as
easily seen in Eq. (\ref{tffardrip-rn}). Freely falling
temperatures for $b<0.5$ can be drawn beyond the horizon (over $p$
and $q$ in the figure) by freely falling observers who are
described by the PG coordinates, while fiducial temperatures
described by the original RN coordinates cannot be drawn due to
divergences at the horizon.}
 \label{fig3}
\end{figure*}

In Fig. \ref{fig2}, we have drawn squared freely falling
temperatures in a drip frame, which can be rewritten in terms of a
dimensionless variable $x=r_+/r$ and a parameter $b=r_-/r_+$ as
 \be \label{tffardrip-rn}
 T_{\rm DF}=\frac{1}{4\pi r^2_+}
            \sqrt{\frac{(1-b)^2(1+x+x^2+x^3)-4bx^4(1-bx)}{1-bx}}.
 \ee
Here, the event horizon of $r=r_+$ ($r=\infty$) corresponds to
$x=1$ $(x=0)$ and the parameter $b$ lies between $0\le b\le 1$.
Note that at the horizon ($x=1$), this is reduced to
 \be
  T_{\rm DF}=\frac{\sqrt{1-2b}}{2\pi r^2_+}.
 \ee
Thus, as seen in Fig. \ref{fig2}, when $b\le 0.5$, squared freely
falling temperatures are finitely well-defined in the whole range
of $x$. On the other hand, when $b>0.5$, there is a region with no
thermal radiation surrounding the horizon, where $T_{\rm DF}$
becomes imaginary. These are exactly the same with the ones in the
BT's work \cite{Brynjolfsson:2008uc}. However, as stated before,
the freely falling temperatures in the PG coordinates are defined
beyond the horizon, while they are not in the original RN
spacetime coordinates. To understand this situation better, we
have drawn freely falling temperatures in the unit of the Hawking
temperature in a drip frame in Fig. \ref{fig3}. Then, one can see
that the freely falling temperatures in a drip frame, which are
obtained from the PG spacetimes, are increased, reached at a peak,
and then decreased to zero as the observers fall into the
singularity after passing through the horizon. Note that if they
are obtained from the original RN coordinates having the spherical
symmetric spacetime (\ref{metric-rn}), freely falling temperatures
are ended at finite values, such as $p$ and $q$, at the horizon in
Figs. \ref{fig2} and \ref{fig3}. Finally, in Fig. \ref{fig3}, the
regions of no thermal radiation are in between the $r=0$
singularity and the positions where the freely falling
temperatures vanish.

(b) A rain frame: for an observer who is freely falling from rest
at the asymptotic infinity, one has $E=1$. Then, the equations of
motion are reduced to
 \bea\label{eomr0rn-rf}
 \frac{dt}{d\tau}&=&1,\nonumber\\
 \frac{dr}{d\tau}&=&-\left(\frac{2m}{r}-\frac{q^2}{r^2}\right)^{1/2}.
 \eea
Again, by exploiting the embedding coordinates in Eq.
(\ref{gems-rngp}), one can obtain a freely falling acceleration
$\bar{a}_{7}$
 \be
 \bar{a}^2_{7}=\frac{(r_+-r_-)^2}{4r^4_+}.
 \ee
Thus, the freely falling temperature at rest at the asymptotic
infinity is
 \be
 T_{\rm RF}=\frac{r_+-r_-}{4\pi r^2_+},
 \ee
which is exactly the same with the Hawking temperature.

(c) A hail frame: if an observer starts to fall with a velocity at
the asymptotic infinity as
 \be
 \frac{dr}{dt}=-v_\infty
 \ee
one has $E=(1-v^2_\infty)^{-1/2}$. Then, the equations of motion
are reduced to
 \bea\label{eomr0rn-hf}
 \frac{dt}{d\tau}&=&\frac{(1-v^2_\infty)^{-1/2}
           -\sqrt{\left(\frac{2m}{r}-\frac{q^2}{r^2}\right)
                \left(\frac{v^2_\infty}{1-v^2_\infty}+\frac{2m}{r}-\frac{q^2}{r^2}\right)}}
           {1-\frac{2m}{r}+\frac{q^2}{r^2}},\nonumber\\
 \frac{dr}{d\tau}&=&-\left(\frac{v^2_\infty}{1-v^2_\infty}+\frac{2m}{r}-\frac{q^2}{r^2}\right)^{1/2}.
 \eea
In this case, one can obtain an acceleration in the hail frame
$\bar{a}_{7}$ by making use of the embedding coordinates in Eq.
(\ref{gems-rngp}) as
 \be
 \bar{a}^2_{7}=\frac{(r_+-r_-)^2}{4r^4_+(1-v^2_\infty)^2}.
 \ee
Thus, the hail frame temperature  is obtained as
 \be
 T_{\rm HF}=\frac{r_+-r_-}{4\pi r^2_+(1-v^2_\infty)}.
 \ee
As in the case of the Schwarzschild black hole in the PG
spacetime, when $v_\infty=0$, $T_{\rm HF}$ reduces to $T_{\rm
RF}$. Finally, the freely falling temperature in the hail frame
can also be rewritten as
 \be
T_{\rm HF}=\frac{E^2(r_+-r_-)}{4\pi r_H}
 \ee
in terms of the energy per unit mass.

\section{Discussion}

In summary, we have globally embedded the (3+1)-dimensional
Schwarzschild and RN black holes in the PG spacetimes into (5+1)-
and (5+2)-dimensional flat ones, respectively, which were made by
the introduction of the PG time. The advantage of introducing the
PG time in the GEMS embeddings is that hypersurfaces constructed
by $z^2$, $z^3$ and $z^4$ either for the Schwarzschild and the RN
black holes remain flat spacelike as in the original spacetimes.
As a result, making use of the embedding coordinates, we have
directly obtained the Hawking, Unruh, and freely falling
temperatures in the drip, rain, and hail frames. Moreover, we have
shown that the Hawking effect for a fiducial observer in a curved
spacetime is exactly equal to the Unruh effect for a uniformly
accelerated observer in higher-dimensionally embedded flat
spacetimes.

On the other hand, the Hawking, Unruh, and freely falling
temperatures in the drip frame in the Schwarzschild and RN black
holes in the PG spacetime have been shown to be well-defined to
the $r=0$ singularity by passing through the event horizon, while
to the event horizon they are exactly the same  as the ones
obtained from the previous GEMS embeddings of the spherically
symmetric Schwarzschild and RN spacetimes. We have also shown that
the freely falling temperatures in the rain frame, which are seen
by an observer who is released freely from the asymptotic
infinity, are nothing but the Hawking ones.

Finally, it seems appropriate to comment that we have shown the
equivalence of the GEMS embeddings of  the spacetimes with the
diagonal and off-diagonal terms in metrics. Therefore, through
further investigation, it would be very interesting to construct
and analyze the GEMS embeddings of the more realistic Kerr, or
rotating BTZ holes by using inverse PG type transformations.

\acknowledgments{S. T. H. was supported by Basic Science Research
Program through the National Research Foundation of Korea funded
by the Ministry of Education, NRF-2019R1I1A1A01058449. Y. W. K.
was supported by the National Research Foundation of Korea (NRF)
grant funded by the Korea government (MSIT) (No.
2020R1H1A2102242). }


\end{document}